\documentclass[3p,times]{elsarticle}

\usepackage{ecrc}
\usepackage{hyperref}

\volume{00}

\firstpage{1}

\journalname{Nuclear Physics A}

\runauth{I. Lakomov for the ALICE Collaboration}

\jid{Nucl.Phys.A}

\jnltitlelogo{Nuclear Physics A}

\CopyrightLine{2014}{Published by Elsevier Ltd.}

\newcommand{\jpsi}{J/$\psi~$}
\newcommand{\jpsino}{J/$\psi$}

\newcommand{\pt}{$p_{ {\mathrm T} }~$}
\newcommand{\ptno}{$p_{ {\mathrm T} }$}

\newcommand{\sqrtsnn}{$\sqrt{s_{\rm NN}}$~}
\newcommand{\gevc}{GeV/$c$~}

\newcommand{\gevcno}{GeV/$c$}

\newcommand{\pp}{\mbox{pp}~}

\newcommand{\cc}{$c\bar{c}$~}

\newcommand{\meanpt}{$\langle p_{ {\mathrm T} } \rangle$~}

\newcommand{\meanptsqrno}{$\langle p_{ {\mathrm T} }^{2} \rangle$}
\newcommand{\npart}{$N_{\rm part}$~}

\usepackage{amssymb}

\usepackage{lineno}

\biboptions{sort&compress}

\usepackage[figuresright]{rotating}

\begin{document}

\begin{frontmatter}



\dochead{}

\title{Event activity dependence of inclusive \jpsi production in p-Pb collisions at \sqrtsnn =~5.02~TeV with ALICE at the LHC}


\author{Igor Lakomov (for the ALICE collaboration)\footnote{A list of members of the ALICE Collaboration and acknowledgements can be found at the end of this issue.}}

\address{Institut de Physique Nucl\'eaire d'Orsay, CNRS-IN2P3, Universit\'e Paris-Sud, 91406 Orsay, France}

\begin{abstract}
Studying p-Pb collisions at LHC energies allows a quantitative evaluation of the cold nuclear matter (CNM) effects. Event activity dependence of the \jpsi production is under intense studies both theoretically and experimentally \cite{Adare:2012qf,Arleo:2013zua,McGlinchey:2012bp,Ferreiro:2012sy,Ferreiro:2013pua} and is expected to provide important insights on the influence of CNM. This paper presents the first results on event activity dependence of the inclusive \jpsi production at backward ($-4.46<y_{\rm cms}<-2.96$) and forward ($2.03<y_{\rm cms}<3.53$) rapidity in ALICE for p-Pb collisions at \sqrtsnn =~5.02~TeV.
\end{abstract}


\end{frontmatter}


\section{Introduction}
\label{}
Heavy-ion collisions are expected to produce sufficiently high energy densities to form a deconfined state of quarks and gluons, referred to as Quark-Gluon Plasma (QGP). One of the signatures of its formation is the so-called ``\jpsi melting'' predicted by Matsui and Satz \cite{Matsui:1986dk}, which results in a suppression of \jpsi production. However in heavy-ion collisions, the effects related to CNM should also be considered in order to correctly interpret the Pb-Pb results. In order to disentangle hot (related to the QGP formation) and CNM effects in heavy-ion collisions, the latter should be quantified precisely. This can be done by studying the nucleon-nucleus collisions where CNM effects play the main role. The CNM effects are usually classified into three groups, depending on the time when they take place, with respect to the time of the \cc pair formation: initial state effects, final state effects and coherent parton energy loss.
{\bf Initial state effects} occur before the \cc pair formation (e.g. gluon shadowing \cite{Ferreiro:2012sy,Ferreiro:2013pua,Albacete:2013ei} or saturation \cite{Kharzeev:2012py}).
{\bf Final state effects} occur after the \cc pair production (e.g. nuclear absorption of \jpsi pre-resonant state).
{\bf Coherent parton energy loss} is neither a pure initial nor a final state effect since it arises from an effect of the interference between gluons radiated by the incoming partons and the \cc pair propagating through the nuclear medium.

In early 2013, the LHC provided p-Pb collisions at \sqrtsnn =~5.02~TeV. The first results on the \jpsi production studies in such collisions were published by the ALICE \cite{Abelev:2013yxa} and LHCb \cite{Aaij:2013zxa} collaborations. A suppression of the~\jpsi production with respect to pp collisions is measured at forward rapidity ($y$) while at backward $y$\ no suppression is found. A~fair agreement is seen with models including nuclear shadowing with the EPS09 parameterizations~\cite{Albacete:2013ei,Ferreiro:2013pua}, and with models including a coherent parton energy loss contribution~\cite{Arleo:2012hn}. For deeper understanding of the CNM effects differential studies are needed. Studying the \jpsi production as a function of the collision centrality will constrain~better CNM effects. The effects are expected to be larger in the most central than in the most peripheral collisions. The latter are expected to be very similar to pp collisions. The collision centrality is determined experimentally~in p-Pb collisions thanks to a Monte-Carlo (MC) generator based on the Glauber model \cite{Glauber:1955qq,Franco:1965wi} that is fitted to the charged-particle multiplicity measurements. The energy measurement of the spectator nucleons can also be associated to a MC generator based on the Slow Nucleon Model (SNM) \cite{Sikler:2003ef}. Due to the bias, discussed below, in the centrality determination, the results depend on the centrality estimator. Therefore, we will use the words ``event activity'' instead of ``centrality''.
\section{Analysis}
ALICE \cite{Abelev:2014ffa} is one of the four large experiments at the LHC. It is dedicated to the study of QGP properties in heavy-ion collisions. \jpsi production can be measured down to zero transverse momentum \pt in two dilepton decay channels: dimuon at forward and dielectron at mid-rapidity. This analysis is based on the dimuon decay channel measured by the ALICE muon spectrometer in $-4<\eta<-2.5$. Due to the beam-energy asymmetry, the nucleon-nucleon center-of-mass (cms) system is shifted with respect to the laboratory frame towards the proton beam direction by $\Delta y = 0.465$. The data were taken with two beam configurations, obtained by inverting the proton and lead beam directions. In the muon spectrometer two rapidity ranges have then been investigated: the backward ($-4.46<y_{\rm cms}<-2.96$) and the forward ($2.03<y_{\rm cms}<3.53$) rapidity. The minimum bias (MB) trigger was defined as the coincidence in the signal of the two scintillator hodoscopes (VZERO) located at both sides of the interaction point. The dimuon trigger used for this analysis, required a coincidence of the MB collision with two opposite-sign muon tracks triggered by the muon trigger chambers. These dimuons are required to have a rapidity $2.5<y^{\mu\mu}_{\rm lab}<4$\ constrained by the muon spectrometer acceptance. The requirement $-4<\eta_{\mu}<-2.5$\ is used to reject single muons at the edge of the acceptance of the spectrometer. To remove tracks crossing the thicker part of the absorber an additional cut was made on $R_{\rm abs}$, the radial transverse position of the muon tracks at the end of the absorber: $17.6<R_{\rm abs}<89.5$~cm. ALICE measures the inclusive \jpsi production while typically the theoretical models deal with prompt \jpsino. The fraction of \jpsi from B decays in the inclusive sample in p-Pb is similar to that in pp collisions ($\sim$10\%) \cite{Aaij:2013zxa}. This effect is well below the quoted experimental uncertainties ($>$10\%). Thus, the ALICE results can be safely compared to theoretical predictions~\cite{Abelev:2013yxa}.

In order to extract the raw \jpsi yield, the opposite-sign dimuon invariant mass distribution is fitted using a superposition of a background and a signal function. The number of \jpsi and its associated statistical and systematic uncertainties are evaluated from different fits varying background and signal functions and the fitting range. Pure \jpsino\ MC simulations tuned on the kinematic distributions of data are used to extract the acceptance-times-efficiency ($A \times \epsilon$).

The centrality of the collision is usually determined considering an experimental observable with a monotonic dependence on the centrality, e.g. the particle multiplicity, or the transverse energy in a certain pseudo-rapidity range. In p-Pb collisions, in contrast to Pb-Pb, the large fluctuations in a much lower particle multiplicity environment, together with the small range of participants, generate a dynamical bias on the centrality determination based on particle multiplicity. In this analysis, the neutron energy distributions in the Zero Degree Calorimeters (ZDC), installed at 112.5~m from the interaction point, were used for the centrality determination since they are less sensitive to this dynamical bias \cite{Alberica:2014qm}. The event activity classes are then determined from the fit to the ZDC neutron (ZN) energy distribution. The class 0-5\% is not presented since this interval is contaminated by pile-up events (events with two or more inelastic p-Pb collisions). An assumption on the mid-rapidity scaling of the charged hadron multiplicity ${\rm d}N_{\rm ch}/{\rm d}\eta$ with \npart (number of participating nucleons) allowed to determine the number of binary nucleon-nucleon collisions $N_{\rm coll}^{\rm mult}$\ and the nuclear thickness function $T_{\rm pPb}^{\rm mult}$\ in each event activity class. The nuclear modification factor is referred to as $Q_{\rm pPb}^{{\rm mult},\,i}$\ when evaluated in a given event activity class $i$, due to the possible bias in the determination of $N_{\rm coll}^{\rm mult}$\ (or $T_{\rm pPb}^{\rm mult}$). It is not necessarily equal to 1 in the absence of nuclear effects. Therefore, the $Q_{\rm pPb}^{{\rm mult},\,i}$\ is defined as:
\begin{equation}\label{eq:QpPb}
Q_{\rm pPb}^{{\rm mult},\,i} = \frac{Y_{\rm pPb}^{i}}{\left<T_{\rm pPb}^{{\rm mult},\,i}\right>\times\sigma_{\rm pp}^{{\rm J/\psi}\rightarrow \mu^{+}\mu^{-}}},
\end{equation}
where $Y_{\rm pPb}^{i}$ is the measured \jpsi invariant yield, $\sigma_{\rm pp}^{{\rm J/\psi}\rightarrow \mu^{+}\mu^{-}}$\ is the \jpsino\ \pp cross-section at the same energy. Due to the absence of pp data at \sqrtsnn=~5.02~TeV, $\sigma_{\rm pp}^{{\rm J/\psi}\rightarrow \mu^{+}\mu^{-}}$\ is interpolated from the ALICE pp results at 2.76 and 7~TeV \cite{ALICE:2013spa}.

\section{Results}
\begin{figure}[h!]
\begin{minipage}{19pc}
\includegraphics[keepaspectratio, width=1.\linewidth]{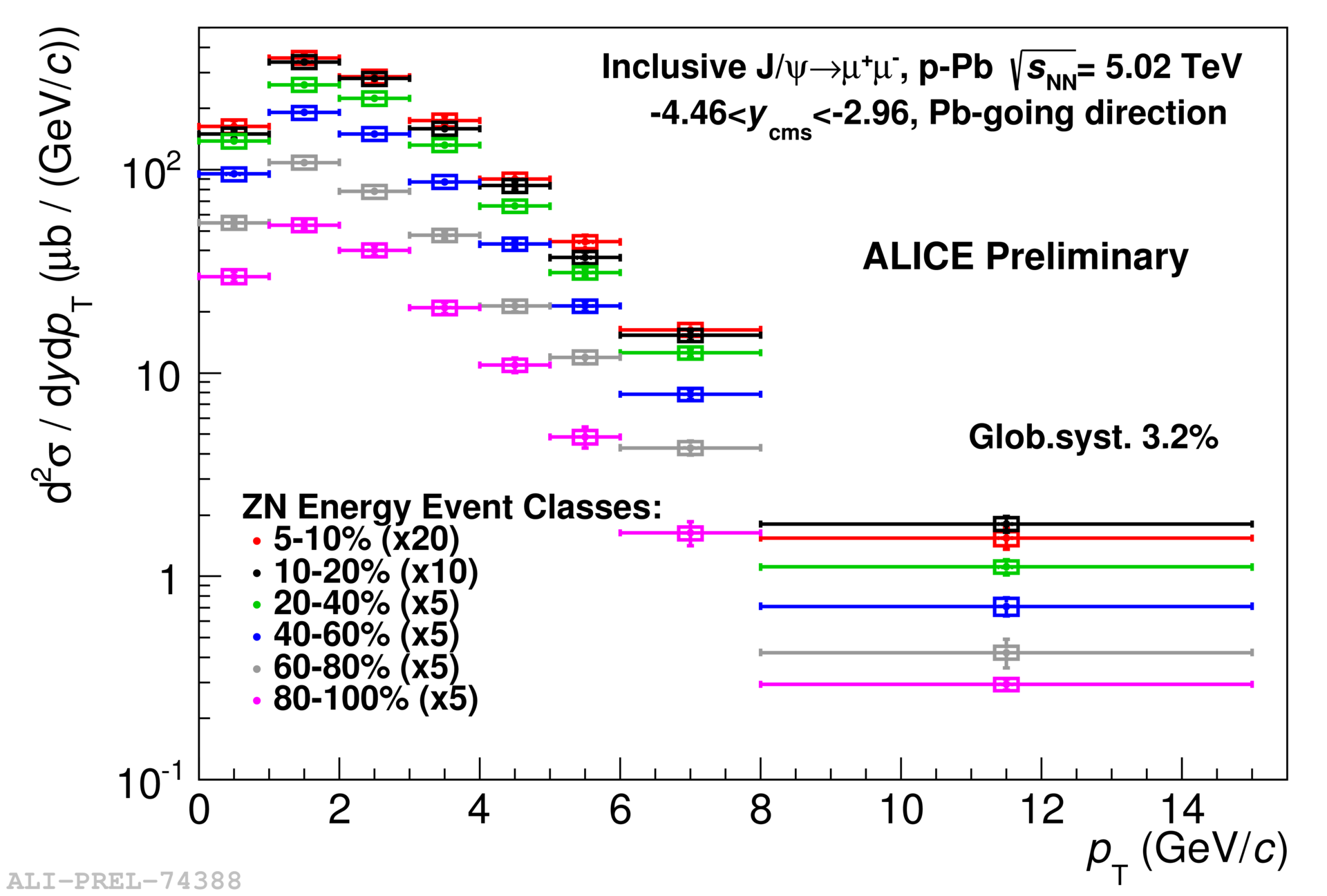}
\end{minipage}
\begin{minipage}{19pc}
\includegraphics[keepaspectratio, width=1.\linewidth]{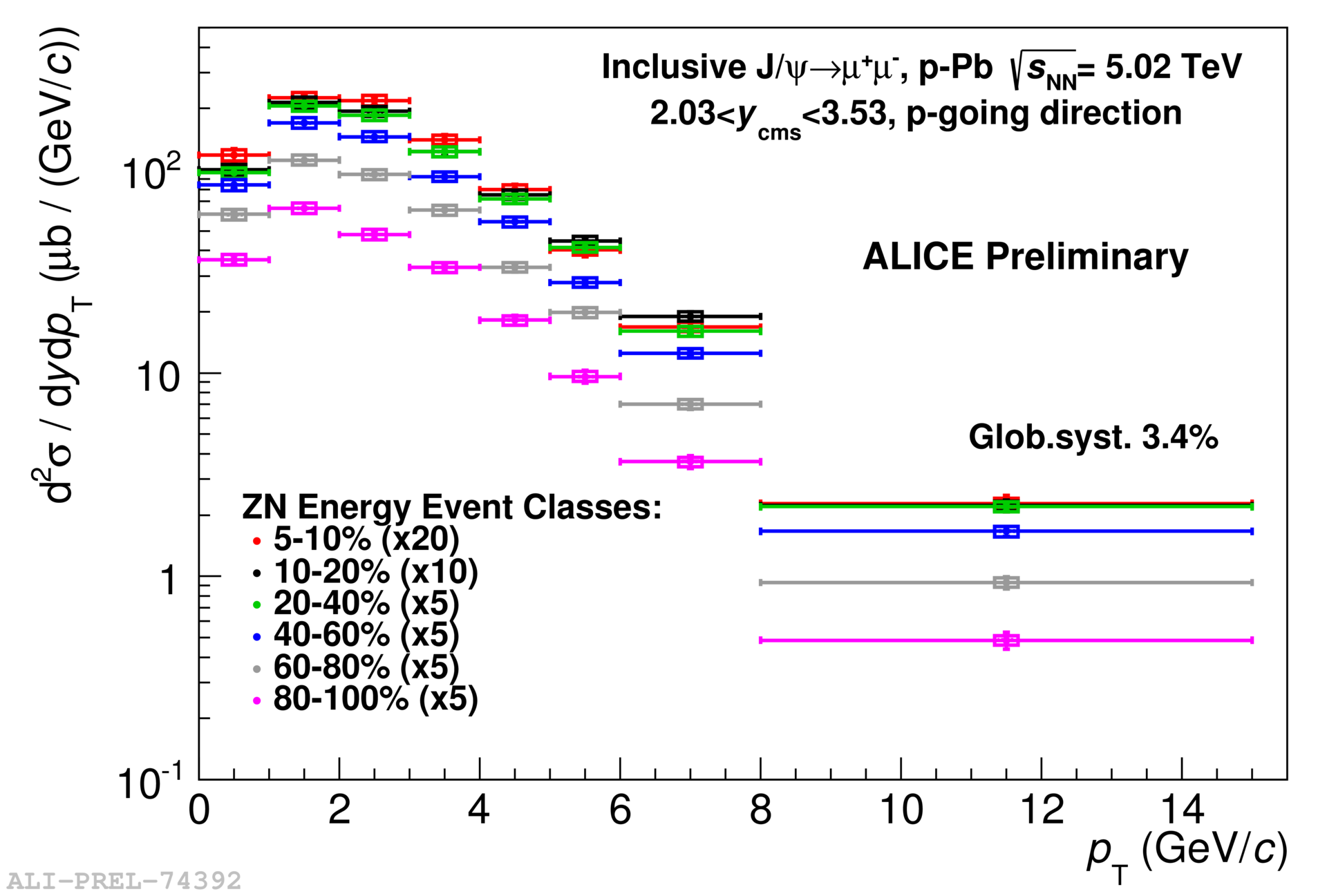}
\end{minipage}
\caption{The \ptno-differential \jpsi cross-sections at backward ({\it left}) and forward ({\it right}) $y$\ in p-Pb collisions for different event activity intervals. The bars are the statistical uncertainties, the open boxes are the uncorrelated and the shaded boxes are the partially correlated systematic uncertainties.}\label{fig:sigmavspt}
\end{figure}

\begin{figure}[h!]
\begin{minipage}{19pc}
\includegraphics[keepaspectratio, width=1.\linewidth]{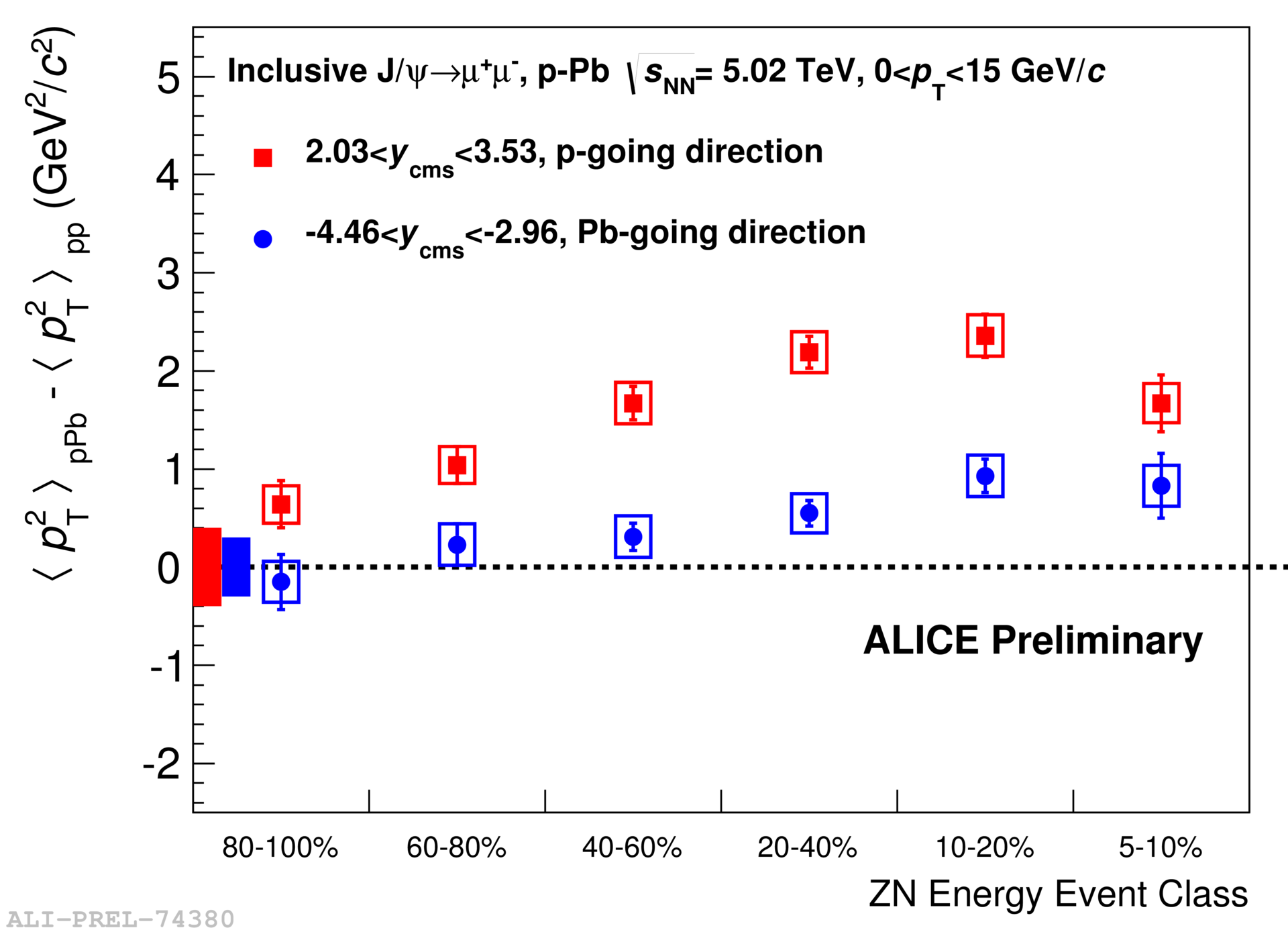}
\end{minipage}
\begin{minipage}{19pc}
\includegraphics[keepaspectratio, width=1.\linewidth]{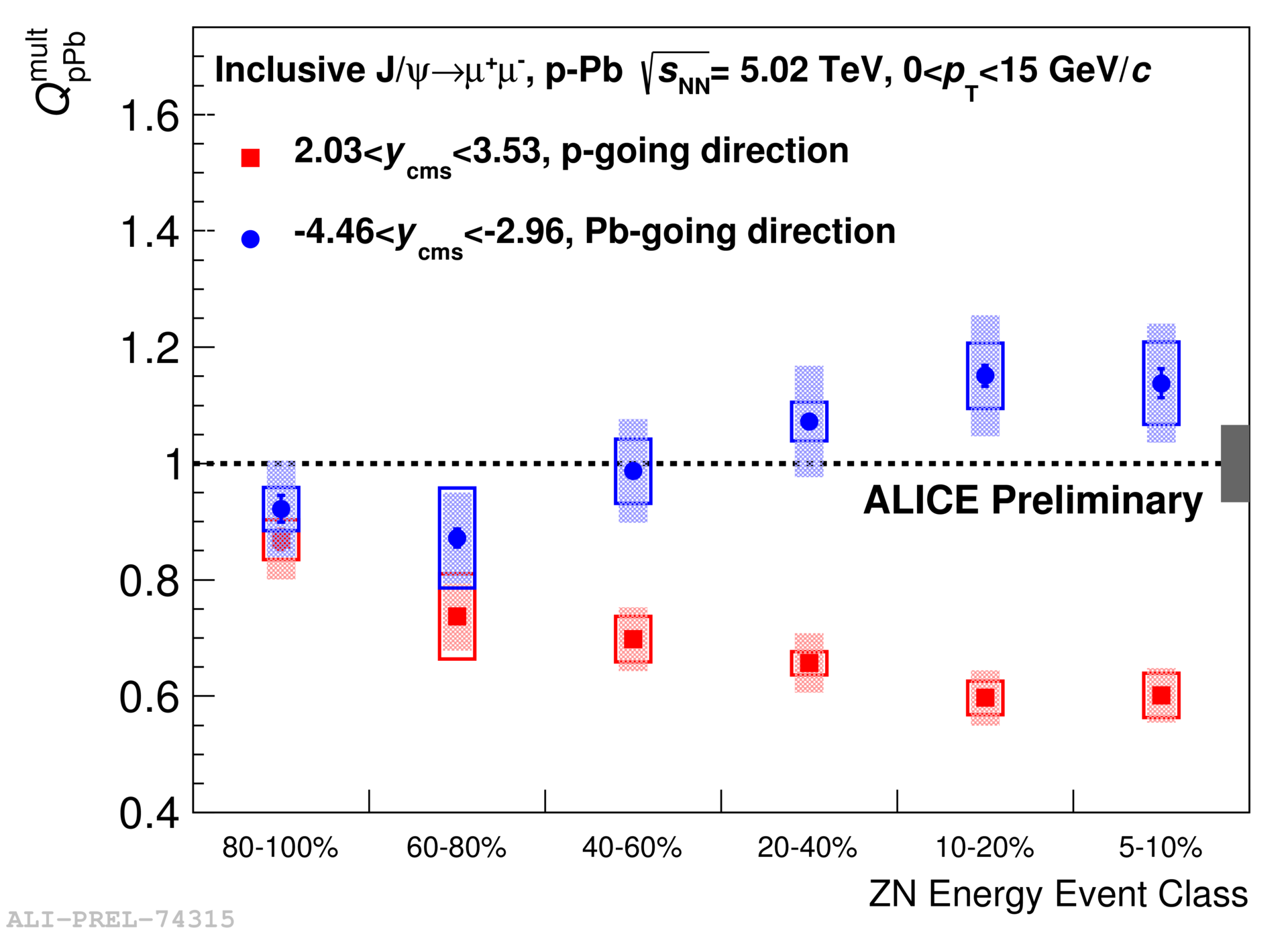}
\end{minipage}
\caption{{\it Left}: Event activity dependence of $\Delta$\meanptsqrno$^{{\rm J}/\psi}_{\rm pPb} \equiv~$\meanptsqrno$^{{\rm J}/\psi}_{\rm pPb}-~$\meanptsqrno$^{{\rm J}/\psi}_{\rm pp}$. The boxes around zero represent the uncertainties of \meanptsqrno$^{{\rm J}/\psi}_{\rm pp}$. {\it Right}:~Event activity dependence of $Q_{\rm pPb}^{\rm mult}$\ integrated over \ptno. The box around unity represents the fully correlated uncertainties.}
\label{fig:Meanptandpt2}
\end{figure}

\begin{figure}[h!]
\begin{minipage}{19pc}
\includegraphics[keepaspectratio, width=1.\linewidth]{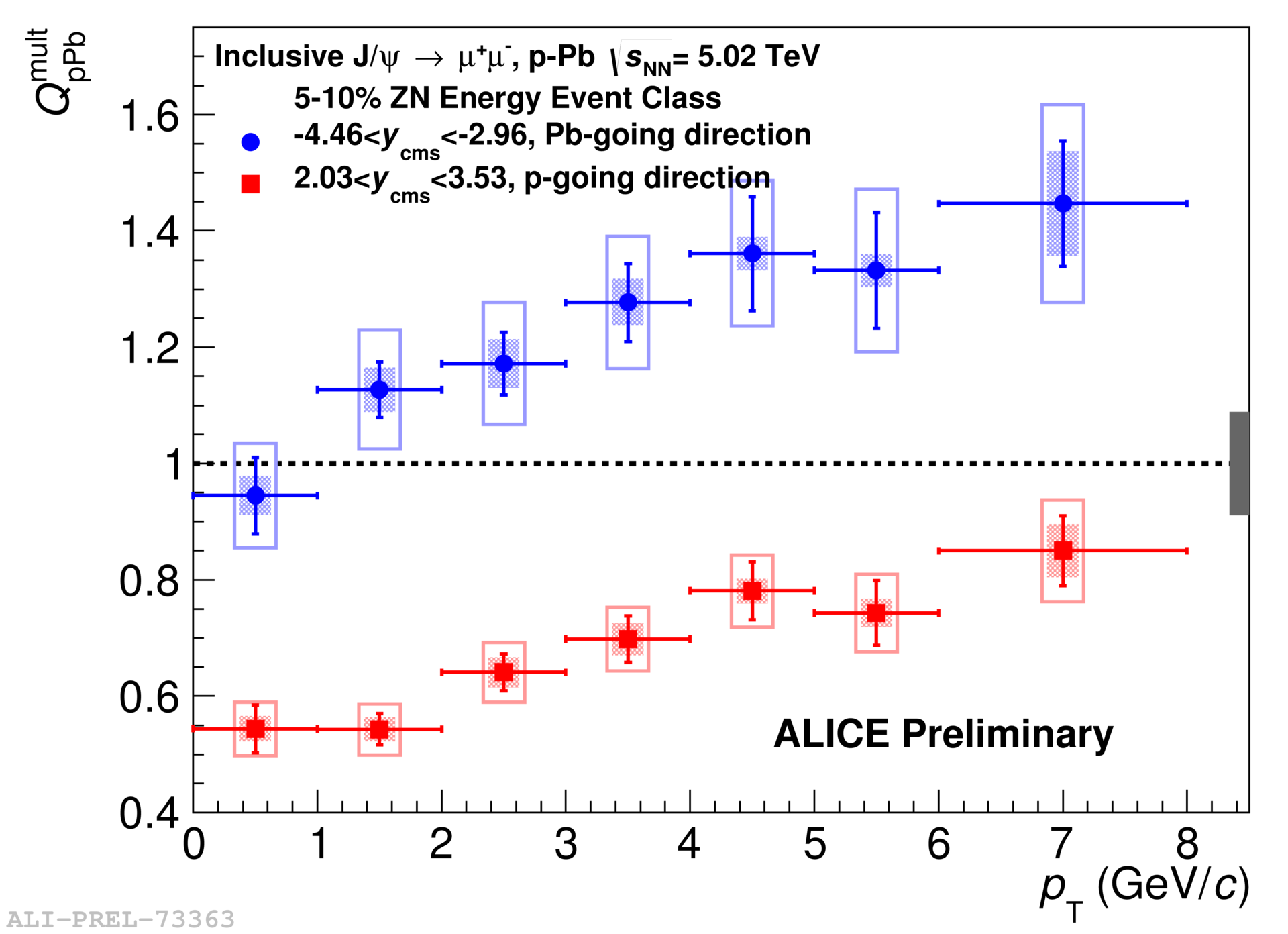}
\end{minipage}
\begin{minipage}{19pc}
\includegraphics[keepaspectratio, width=1.\linewidth]{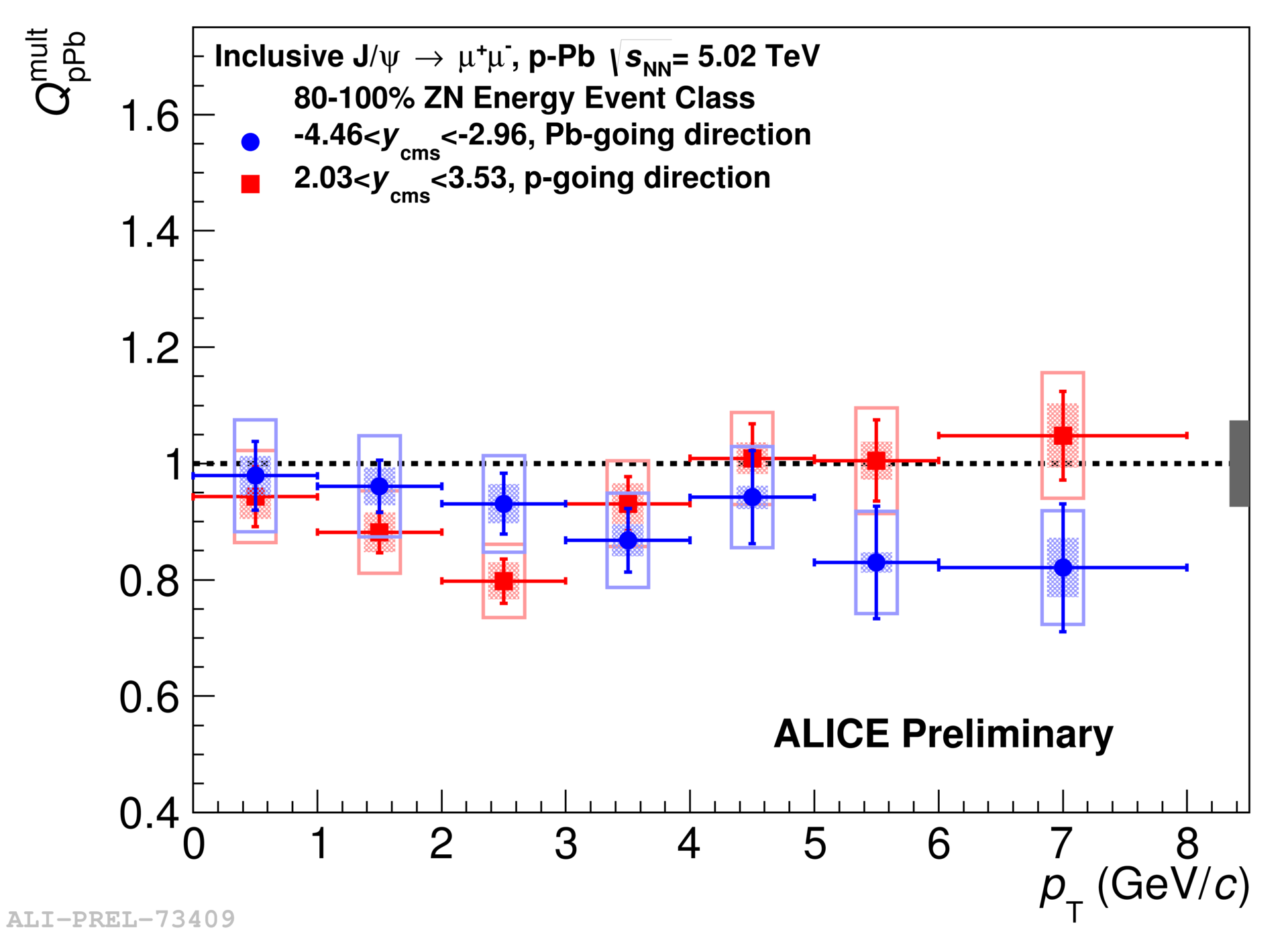}
\end{minipage}
\caption{The \pt dependence of $Q_{\rm pPb}^{\rm mult}$\ at large ({\it left}) and small ({\it right}) event activity classes. The bars are the statistical uncertainties, the open boxes are the uncorrelated, the shaded boxes are the partially correlated and the boxes around unity are the fully correlated systematic uncertainties.}
\label{fig:QpPbvspt}
\end{figure}

Figure~\ref{fig:sigmavspt} shows the \ptno-differential \jpsi cross-sections in p-Pb collisions in different event activity intervals for backward (left) and forward (right) rapidity. For a better visualization, arbitrary scaling factors are applied to the distributions for different event activity classes (shown in brackets in the figures).
The \pt distributions were fitted with a function $f(p_{\rm T})=C\frac{p_{\rm T}}{\left(1+\left(p_{\rm T}/p_{0}\right)^{2}\right)^{n}}$ in p-Pb collisions in different event activity classes and in pp collisions by using the interpolated distributions. The first and second momenta of $f(p_{\rm T})$, evaluated for $0<p_{\rm T}<15$~\gevc determines the \jpsi \meanpt and \meanptsqrno, respectively. The event activity dependence of $\Delta$\meanptsqrno$^{{\rm J}/\psi}_{\rm pPb} \equiv~$\meanptsqrno$^{{\rm J}/\psi}_{\rm pPb}-~$\meanptsqrno$^{{\rm J}/\psi}_{\rm pp}$\ is shown in the left panel of Fig.\ref{fig:Meanptandpt2}. The \jpsi production in p-Pb has a harder \pt distribution at forward than at backward $y$ for the full event activity range. At small event activity (80-100\%) at backward $y$\ \meanptsqrno$^{{\rm J}/\psi}_{\rm pPb}\approx$\meanptsqrno$^{{\rm J}/\psi}_{\rm pp}$. The shape and the amplitude of $\Delta$\meanptsqrno$^{{\rm J}/\psi}_{\rm pPb}$\ at backward $y$\ are similar to those observed in d-Au collisions at \sqrtsnn =~200~GeV by PHENIX~\cite{Adare:2012qf} (not shown). At forward $y$\ the ALICE results show a higher amplitude.
The right panel of Fig.~\ref{fig:Meanptandpt2} shows the event activity dependence of $Q_{\rm pPb}^{\rm mult}$. A large difference is seen between the event activity dependence at backward and forward $y$. At backward $y$, $Q_{\rm pPb}^{\rm mult}$\ is consistent with unity for the full event activity range. At forward $y$\ the suppression of \jpsino\ production increases with the event activity up to 45\%.
In Fig.~\ref{fig:QpPbvspt} the \pt dependence of $Q_{\rm pPb}^{\rm mult}$\ is shown for large (5-10\%, left) and small (80-100\%, right) event activity. At large event activity $Q_{\rm pPb}^{\rm mult}$\ increases with \pt both at backward and forward $y$. At backward $y$ it is higher than~1 and reaches 1.45 at \pt=~7~\gevcno. It is qualitatively described by the predictions from \cite{Arleo:2013zua}. At small event activity $Q_{\rm pPb}^{\rm mult}$\ is consistent with unity for both backward and forward $y$\ in the full \pt range. It agrees with theoretical expectations \cite{Arleo:2013zua,Ferreiro:2012sy,Ferreiro:2013pua}. As discussed before, a bias in the determination of $T_{\rm pPb}^{\rm mult}$\ can affect the $Q_{\rm pPb}^{\rm mult}$\, but the ratio of forward and backward~$y$\ results should not be affected by such a bias. This ratio for low \pt$\lesssim 1$~\gevcno\ changes from 0.54 to about 0.95 when going from large to small event activity.
\section{Conclusions}
Event activity dependence of the \jpsi production in p-Pb collisions at \sqrtsnn =~5.02~TeV was studied in ALICE using the ZDC as event activity estimator. \jpsi production has a harder \pt distribution at forward than at backward $y$. At small event activity (80-100\%) at backward~$y$, \meanptsqrno$^{{\rm J}/\psi}_{\rm pPb}\approx$\meanptsqrno$^{{\rm J}/\psi}_{\rm pp}$. According to this estimator, at small event activity no nuclear effects are seen while at large event activity they are large: $Q_{\rm pPb}^{\rm mult}\simeq 1.45$\ at \pt$\sim$~7~\gevcno\ at backward $y$\ and $Q_{\rm pPb}^{\rm mult}\simeq 0.55$\ at \pt$\lesssim 1$~\gevcno\ at forward $y$. The ratio of forward and backward $y$\ results also changes dramatically with the event activity: for low \pt$\lesssim 1$~\gevcno\ it changes from 0.54 to about 0.95 when going from large to small event activity. A quantitative comparison with theoretical models is needed for a better interpretation of these results. 
\bibliographystyle{elsarticle-num}
\bibliography{./Bibliography}

\begin{thebibliography}{10}
\expandafter\ifx\csname url\endcsname\relax
  \def\url#1{\texttt{#1}}\fi
\expandafter\ifx\csname urlprefix\endcsname\relax\def\urlprefix{URL }\fi
\expandafter\ifx\csname href\endcsname\relax
  \def\href#1#2{#2} \def\path#1{#1}\fi

\bibitem{Adare:2012qf}
A.~Adare, et~al., Phys. Rev. C87 (2013) 034904.
\newblock \href {http://arxiv.org/abs/1204.0777} {\path{arXiv:1204.0777}}.

\bibitem{Arleo:2013zua}
F.~Arleo, R.~Kolevatov, S.~Peign\'e, M.~Rustamova, JHEP 1305 (2013) 155.
\newblock \href {http://arxiv.org/abs/1304.0901} {\path{arXiv:1304.0901}}.

\bibitem{McGlinchey:2012bp}
D.~McGlinchey, A.~Frawley, R.~Vogt, Phys. Rev. C87 (2013) 054910.
\newblock \href {http://arxiv.org/abs/1208.2667} {\path{arXiv:1208.2667}}.

\bibitem{Ferreiro:2012sy}
E.~Ferreiro, F.~Fleuret, J.~Lansberg, N.~Matagne, A.~Rakotozafindrabe, Few Body
  Syst. 53 (2012) 27--36.
\newblock \href {http://arxiv.org/abs/1201.5574} {\path{arXiv:1201.5574}}.

\bibitem{Ferreiro:2013pua}
E.~Ferreiro, F.~Fleuret, J.~Lansberg, A.~Rakotozafindrabe, Phys. Rev. C88
  (2013) 047901.
\newblock \href {http://arxiv.org/abs/1305.4569} {\path{arXiv:1305.4569}}.

\bibitem{Matsui:1986dk}
T.~Matsui, H.~Satz, Phys. Lett. B178 (1986) 416.

\bibitem{Albacete:2013ei}
J.~Albacete, N.~Armesto, R.~Baier, G.~Barnafoldi, J.~Barrette, et~al.,
  Int.J.Mod.Phys. E22 (2013) 1330007.
\newblock \href {http://arxiv.org/abs/1301.3395} {\path{arXiv:1301.3395}}.

\bibitem{Kharzeev:2012py}
D.~Kharzeev, E.~Levin, K.~Tuchin, Nucl.Phys. A924 (2014) 47--64.
\newblock \href {http://arxiv.org/abs/1205.1554} {\path{arXiv:1205.1554}}.

\bibitem{Abelev:2013yxa}
B.~B. Abelev, et~al., JHEP 1402 (2014) 073.
\newblock \href {http://arxiv.org/abs/1308.6726} {\path{arXiv:1308.6726}}.

\bibitem{Aaij:2013zxa}
R.~Aaij, et~al., JHEP 1402 (2014) 072.
\newblock \href {http://arxiv.org/abs/1308.6729} {\path{arXiv:1308.6729}}.

\bibitem{Arleo:2012hn}
F.~Arleo, S.~Peign\'e, Phys. Rev. Lett. 109 (2012) 122301.
\newblock \href {http://arxiv.org/abs/1204.4609} {\path{arXiv:1204.4609}}.

\bibitem{Glauber:1955qq}
R.~Glauber, Phys. Rev. 100 (1955) 242--248.

\bibitem{Franco:1965wi}
V.~Franco, R.~Glauber, Phys. Rev. 142 (1966) 1195--1214.

\bibitem{Sikler:2003ef}
F.~{Sikler. }\href {http://arxiv.org/abs/hep-ph/0304065}
  {\path{arXiv:hep-ph/0304065}}.

\bibitem{Abelev:2014ffa}
B.~B. Abelev, {et al. }\href {http://arxiv.org/abs/1402.4476}
  {\path{arXiv:1402.4476}}.

\bibitem{Alberica:2014qm}
A.~{Toia for the ALICE Collaboration}, These proceedings.

\bibitem{ALICE:2013spa}
{ALICE and LHCb collaborations}, ALICE-PUBLIC-2013-002, LHCB-CONF-2013-013-002.

\end{thebibliography}
\end{document}